\documentclass[prl,aps,twocolumn,showpacs]{revtex4-1}

\def\pgatch{$^1$}

\begin{document}
\vspace*{-2in}

\title{
Comment on "Froissart bound on total cross section\\
without unknown constants" by A. Martin and S.M. Roy
}

\author{
Ya.I.~Azimov\pgatch
}

\affiliation{
\pgatch Petersburg Nuclear Physics Institute, 188300 Gatchina, Russia}


\begin{abstract}
Here I explain why critics of my work in the above paper
is inadequate.
\end{abstract}


\maketitle

The recent paper~\cite{MR} criticized my earlier
publication~\cite{az}. Regretfully, presentation of
my work by Martin and Roy looks to be incorrect.
Therefore, it is necessary to clarify the situation.

1. According to Martin and Roy~\cite{MR}, Sec.~2 of my
paper ``is similar to the 1962 and 1963 works of
Martin''~\cite{M62,M63}. The talk~\cite{M62} gives only
a brief presentation of results of the much more detailed
paper~\cite{M63}. The approach used by the both publications
is definitely different from mine~\cite{az}. Martin studies
there only the imaginary part of the scattering amplitude.
As a result, he could derive a bound for the total cross
section (through the optical theorem), but not for differential
cross sections. Contrary to that, I followed to
Froissart~\cite{frois} and have not separated the real
and imaginary parts, which allowed me (again, following
to Froissart) to constrain both total and differential
cross sections.

My approach is, in essence, very similar to Froissart's one.
They differ mainly in the manner of using assumptions on the
asymptotic behavior of the amplitude at the large circles
of the energy and momentum transfer. Froissart assumes them
to provide dispersion relations in energy and momentum transfer,
with a finite number of subtractions. On the other side, I do
not need any dispersion relations at all and is able to use
high energy asymptotics of the amplitude as a separate arbitrary
assumption which admit variations. This allows to analyze role
of various assumptions. More detailed discussion of similarity
and difference between these two approaches can be found in
Ref.\cite{az}.

In any case, my paper~\cite{az} is not ``similar to the 1962
and 1963 works of Martin''~\cite{M62,M63}.

2. In his various publications, starting from the papers~\cite{MNC42A},
Martin has used axiomatics of local field theory. In Introduction to
my paper~\cite{az}, I noted that QCD might appear not corresponding to
such axiomatics, since quarks and/or gluons do not have asymptotic
states, while hadrons are not local objects (they consist of quarks
and have, therefore, internal structure). Now, Martin and Roy~\cite{MR}
oppose this note, referring to Zimmermann~\cite{Zim}, who ``has shown
that local fields can be associated to composite particles (for
instance, deuterons)''. This is true, of course, but a field theory
expressed trough such local fields may look non-local. For instance,
interactions of deuterons should reveal their form factors. Similarly,
QCD may (and, most probably, should) appear non-local, being expressed
in terms of hadron fields.

Curiously enough, Martin and Roy really support and even expand my note.
Indeed, just after opposing it and reminding of Zimmermann's results, 
they write: ``We postulate that this construction applies to hadrons made 
of quarks. This is not obvious because, in spite of the practical successes
of QCD, nobody knows how to incorporate particles without asymptotic
fields in a field theory.'' Thus, they clearly agree that applicability
of Martin's assumptions to QCD, and to hadron physics,  is not evident
and needs indeed to be specially postulated.

3. As an additional point, Martin and Roy say that they ``do not
use the Froissart-Gribov representation of physical region partial
waves for fixed $s$.'' I should emphasize here, that my approach
does not use this representation as well.

In conclusion, this analysis shows that critics of my paper~\cite{az}
by Martin and Roy~\cite{MR} is inadequate.


\end{document}